\documentclass[12pt]{article}
\usepackage{amsmath}
\usepackage{graphics,graphicx,epsfig}
\usepackage{amssymb}
\usepackage{epstopdf}
\usepackage{sidecap}
\usepackage{appendix}

\def\be{\begin{equation}}
\def\ee{\end{equation}}
\def\bea{\begin{eqnarray}}
\def\eea{\end{eqnarray}}
\def\ten{ {\cal T} }

\begin{document}

\begin{titlepage}
\vfill
\begin{flushright}
ACFI-T16-21
\end{flushright}

\vskip 0.4in

\begin{center}
\baselineskip=16pt
{\Large\bf Genuine Cosmic Hair}
\vskip 0.15in

{\bf  David Kastor${}^{a}$, Sourya Ray${}^{b}$, Jennie Traschen${}^{a}$} 

\vskip 0.4cm
${}^a$ Amherst Center for Fundamental Interactions, Department of Physics\\ University of Massachusetts, Amherst, MA 01003\\

\vskip 0.1in ${}^b$ Instituto de Ciencias F\'{\i}sicas y Matem\'{a}ticas, Universidad Austral de
Chile, Valdivia, Chile\\
\vskip 0.1 in Email: \texttt{kastor@physics.umass.edu, ray@uach.cl, traschen@physics.umass.edu}
\vspace{6pt}
\end{center}
\vskip 0.2in
\par
\begin{center}
{\bf Abstract}
 \end{center}
\begin{quote}

We show that  asymptotically future deSitter (AFdS) spacetimes carry  `genuine' cosmic hair;   
information  that is analogous to the mass and angular momentum of asymptotically flat spacetimes and that characterizes how an AFdS spacetime approaches its asymptotic form.  We define new `cosmological tension' charges associated with future asymptotic spatial translation symmetries, which are analytic continuations of the ADM mass and tensions of asymptotically planar AdS spacetimes, and which measure the leading anisotropic corrections to the isotropic, exponential deSitter expansion rate.
A cosmological Smarr relation, holding for AFdS spacetimes having exact spatial translation symmetry, is derived. This formula
relates cosmological tension, which is evaluated at future infinity, to properties of the cosmology at early times, together with a
 `cosmological volume' contribution that is analogous to the thermodynamic volume of AdS black holes. 
 Smarr relations for different spatial directions  imply that the difference in expansion rates between two directions at late times
 is related in a simple way to their difference at early times.  Hence information about the very early universe can be inferred from cosmic hair, which is potentially observable in a late time deSitter phase.  Cosmological tension charges and related quantities are evaluated for Kasner-deSitter spacetimes, which serve as our primary examples.

\vfill
\vskip 2.mm
\end{quote}
\hfill
\end{titlepage}


\section{Introduction}

A sufficiently long period of inflation is expected to wipe out pre-existing spatial inhomogeneities and anisotropies.  
An important result on the decay of anisotropy was established in \cite{Wald:1983ky}, where it was shown that a homogeneous, initially expanding cosmology with $\Lambda>0$, and matter that satisfies the dominant and strong energy conditions, approaches deSitter spacetime exponentially quickly (with the exception of certain Bianchi type IX spacetimes).  This result is commonly referred to in the literature as the ``cosmic no hair theorem".  However, we will show that this characterization is inaccurate and that asymptotically future deSitter (AFdS) spacetimes, such as those considered in \cite{Wald:1983ky}, in fact, have persistent cosmic hair that preserves a certain amount of information on anisotropy even in the asymptotic regime.

The term ``hair", in the gravitational context, was originally coined for black holes, and refers to information required to characterize an equilibrium state.  Stationary, asymptotically flat black holes in four dimensions have mass, charge and angular momentum.   These quantities are determined by the leading fall-off coefficients of components of the metric and gauge potential in the asymptotic region, by means of the ADM prescription for conserved charges associated with asymptotic symmetries \cite{Arnowitt:1962hi} and Gauss's law.  We can consider such quantities to be ``fundamental hair".  No hair theorems address whether any additional information, beyond this fundamental hair, is also required to characterize black hole hole equilibrium states\footnote{See, for example, references \cite{Herdeiro:2015waa,Sotiriou:2015pka} for recent reviews of both no hair theorems, and examples where black hole hair exists, in Einstein gravity coupled to different types of scalar field theories.}.
The fact that a black hole in equilibrium is characterized only by a small set of physical parameters, which exclude detailed information regarding its formation and evolution, is central to the analogy between black holes and thermodynamic systems that is realized in the laws of black hole mechanics \cite{Bardeen:1973gs}. 

Returning to cosmology, we will show in this paper that AFdS spacetimes, such as those considered in \cite{Wald:1983ky}, are characterized by a type of cosmic hair, which for reasons to be discussed below we refer to as cosmological tension.  Like the mass and angular momentum of an asymptotically flat black hole, the cosmological tension of an AFdS spacetime is determined by the falloff behavior of the spacetime metric in its asymptotic regime, and should therefore be regarded as fundamental cosmic hair\footnote{A real cosmic no hair theorem, which is beyond the scope of our work, should address to what extent cosmological tension entirely characterizes AFdS spacetimes.}.  

We will show below that cosmological tension is an ADM charge, defined at future infinity in an AFdS spacetime.
An ADM charge is generally given by an integral over a codimension-$2$ surface $\partial\Sigma_\infty$, which is the boundary at infinity in an asymptotic region of the spacetime of a codimension-$1$ slice $\Sigma$.  Two key ingredients in the definition of a given charge are choices in the asymptotic region of an asymptotic symmetry generator and the direction of the normal to the 
slice $\Sigma$.  The actual location of the asymptotic region, where falloff conditions are imposed on the metric, represents a third key ingredient.  Although this is taken to be spatial infinity in most familiar examples, the ADM charges for AFdS spacetimes are naturally defined at future infinity.  

Consider the choices for these three key ingredients in various examples.
The ADM mass of an asymptotically flat spacetime is evaluated at spatial infinity  and is defined in terms of the asymptotic time translation symmetry, with the normal to 
$\Sigma$ taken to point in the time direction.  The ADM angular momentum is similarly defined, using the asymptotic azimuthal symmetry.  The ADM spatial tensions defined in \cite{Traschen:2001pb,Townsend:2001rg,Harmark:2004ch},  which will be particularly relevant for our construction, are defined for spacetimes like those describing black strings and $p$-branes that are ``transverse asymptotically flat'', such that the metric falls off to the flat metric only in directions transverse to the brane, and not in spatial directions tangent to the brane worldvolume.  The ADM mass is defined in the usual way for such spacetimes, by choosing the asymptotic time translation symmetry and a slice $\Sigma$ with normal pointing in the time direction.  However, there is also a spatial tension for each direction tangent to the brane worldvolume, defined by choosing both the asymptotic spatial translation symmetry and the normal to the slice 
$\Sigma$ to be in a spatial direction along the brane\footnote{The ADM mass can be thought of, in this scheme, as the tension in the time direction.  Note also that the spatial tension of a $p$-brane spacetime in a given direction is distinct from the ADM momentum of the brane in that direction.  The momentum is defined in terms of the same asymptotic spatial translation vector as the tension, but with the normal to the slice $\Sigma$ pointing in the time direction.}.

Now consider what types of ADM charges can be defined in the future asymptotic region of an AFdS spacetime.
There will no longer be an ADM mass, because the metric depends on time and the time direction is  the direction of approach to the asymptotic region, analogous to the radial direction in asymptotically flat spacetimes. However, there will be ADM tensions in each of the spatial directions, defined in analogy with the spatial tensions of transverse asymptotically flat spacetimes, by taking an asymptotic spatial translation symmetry and a slice $\Sigma$ with normal pointing in the same spatial direction. The difference between the spatial and 
cosmological tensions is that in the cosmology the boundary integral is on $\partial\Sigma_\infty$ at future infinity where the metric falls off in time, whereas 
for spatial tension the boundary is as spatial infinity where the metric falls off in a radial coordinate.
 The cosmological tensions constitute the fundamental cosmic hair for AFdS spacetimes.  We will see, in particular, that although the homogeneous spacetimes considered in  \cite{Wald:1983ky} approach deSitter spacetime exponentially fast, this rate of decay of the perturbations turns out to be precisely what is required to balance
 the exponential growth of the area element in deSitter to
  give finite values for the spatial integrals that yield the cosmological tension charges\footnote{This is analogous to asymptotically flat spacetimes, where the falloff of metric perturbations in the asymptotic region with inverse powers of the radius is compensated for by the growth of the area element of spheres to give finite results for ADM integrals,
  or to the competing effects that give finite charges in AdS.}.

The cosmological tension $\ten_{(y)}$ associated with translation symmetry
in the $y$-direction works out to be the leading order correction to the Hubble expansion rate in the $y$-direction compared to the asymptotic isotropic deSitter expansion rate.
The tensions are not all independent, rather they must sum to zero as a consequence of the field equations in the asymptotic region. This is the analog for AFdS spacetimes of the tracelessness of the boundary stress tensor for asymptotically AdS spacetimes.

We also construct a Smarr relation involving the cosmological tension, which holds for AFdS spacetimes with exact spatial translation symmetry. Recall that in the case of static AdS black holes, the time translation symmetry can be used to derive a Smarr formula \cite{Kastor:2009wy} that relates the ADM mass to 
properties of the black hole horizon, namely $\kappa A$. In AdS the Smarr relation also has a contribution of the form $\Lambda V_{thermo}$, where  the thermodynamic volume is a finite quantity that can be viewed either as the volume ``inside" the black hole, or as the volume of the region of anti-deSitter spacetime displaced by the presence of the black hole.  This viewpoint leads to interpreting the ADM mass parametter of an asymptotically AdS black hole more properly as the spacetime enthalpy.

The cosmological Smarr formula, which we derive,
relates the cosmological tension, defined on the late time boundary, to properties of the cosmology at an early time, namely 
the Hubble expansion in the symmetry direction integrated over the early time surface.
Perhaps surprisingly, when the early time is taken to be the big bang, this term remains finite for a range of 
interesting cosmologies. 
The Smarr formula also receives a contribution from the stress-energy present in the spacetime. 
In the case of 
Einstein gravity with a cosmological constant $\Lambda$, and no additional matter fields, this contribution 
has the form $\Lambda V_{cosmo} $, where the cosmological volume can be viewed either
 as the spacetime volume ``before" the big bang singularity, or as the region of deSitter spacetime cutoff by the introduction of the singularity\footnote{Although we have not done this, it would be interesting to develop a notion of cosmological enthalpy in analogy with the AdS black hole case \cite{Kastor:2009wy}.}.  If there is exact translation symmetry in more than one spatial direction, {\it e.g.} $y$ and $x$, then the corresponding Smarr formulas can be combined to show that the difference of the leading anisotropic expansion rates in the two directions, given by $\ten_{(y)}- \ten_{(x)}$,
 is equal to the integrated difference in the expansion rates at the early time boundary.
 This result shows that information from the very early universe can potentially be inferred from cosmic hair measured in the late time deSitter phase.
 
The plan of the paper is as follows.  In section (\ref{kasnersec}) we introduce the Kasner-deSitter spacetimes, a family of AFdS solutions to Einstein gravity with a positive cosmological constant, which we will use as examples to illustrate the general constructions in subsequent sections.  In section (\ref{admsection}) we define ADM cosmological tension for AFdS spacetimes, evaluated in the late time asymptotic limit, using the Hamiltonian framework, and compute the cosmological tensions for Kasner-deSitter spacetimes.
In section (\ref{komarsection}) we present a Komar version of cosmological tension in cosmologies that have exact spatial translation symmetry.   These enter
the  cosmological Smarr relations, which are derived for general homogeneous but anisotropic cosmologies. In section (\ref{smarrsection}) we tailor the Smarr relation to AFdS spacetimes.  This requires a modification of the Komar construction, similar to that made for asymptotically AdS black holes \cite{Kastor:2009wy}, and leads to the concept of `cosmological volume' for AFdS spacetimes, in analogy with the thermodynamic volume of asymptotically AdS black holes \cite{Kastor:2009wy}.  In section (\ref{conclusion}) we briefly summarize our results and point to some directions for future work.  Appendix (\ref{cosmovolume}) contains a more general treatment of cosmological volume.

\section{Kasner-deSitter spacetimes}\label{kasnersec}

In order to illustrate cosmological tension, it is useful to have an example in mind.  For this purpose, we introduce the Kasner-deSitter spacetimes, which have appeared multiple times in the literature\footnote{See, for example, the recent discussion of a primordial anisotropic stage of inflation in \cite{Blanco-Pillado:2015dfa}.}.  These are solutions to Einstein gravity with a positive cosmological constant,
\begin{equation}\label{action}
S=\int d^D x\sqrt{-g}(R-2\Lambda)
\end{equation}
which are asymptotic at early times to Kasner spacetimes and at late times to deSitter spacetime.  They are anisotropic, but homogeneous spacetimes and therefore fall into the class considered in \cite{Wald:1983ky}.
Recall that Kasner spacetimes are homogeneous, but anisotropic cosmological solutions to the vacuum Einstein equations. 
 In $D=n+1$ dimensions Kasner spacetimes are given by
\begin{equation}\label{kasner}
ds^2=-dt^2 +\sum_{k=1}^n \left({t\over t_0}\right)^{2p_k} (dx^k)^2
\end{equation}
where the exponents satisfy the Kasner constraints
\begin{equation}\label{kasnerconstraints}
\sum_{k=1}^n p_k = 1 =\sum_{k=1}^n (p_k)^2
\end{equation}
The Kasner solutions have a curvature singularity at $t=t_0$, with the exception of the case that a single exponent {\it e.g.} $p_1=1$ with the remaining exponents vanishing, which gives Minkowski spacetime in Milne coordinates.

The Kasner-deSitter solutions to (\ref{action}) are given in terms of a proper time coordinate by
\begin{equation}\label{cosmometric}
ds^2 = -dt^2 +{T_0^2\over l^2}\cosh^{4/n}({nt\over 2l})\left\{
\sum_{k=1}^n \tanh^{2p_k}({nt\over 2l})(dx^k)^2\right\}
\end{equation}
where we have parameterized the cosmological constant in terms of a deSitter curvature radius $l$ according to
\begin{equation}\label{desitter}
\Lambda = {n(n-1)\over 2l^2}
\end{equation}
The exponents $p_k$ again satisfy the Kasner constraints (\ref{constraints}).   The dimensionless factor $T_0^2/l^2$ is inserted for later convenience.  
In the limit of early times $nt/2l\ll 1$ the Kasner-deSitter spacetimes (\ref{cosmometric}) reduce, after a rescaling of the spatial coordinates, to the Kasner spacetimes (\ref{kasner}).
In the late time limit 
$nt/2L\gg 1$, one finds an exponential decay towards an asymptotically deSitter metric given by
\begin{equation}
ds^2 \simeq -dt^2 +e^{2Ht}\sum_{n=1}^k(1+ {4(1+np_k)\over n}e^{-nHt})(d\tilde x^k)^2
\end{equation}
where $H=1/l$ is the Hubble constant and the spatial coordinates have been rescaled by a constant factor relative to (\ref{cosmometric}).  This asymptotic behavior of the metric accords with the results of \cite{Wald:1983ky}.  However, note also that the exponentially decaying terms carry information about the early state of the spacetime through the Kasner exponents $p_k$.  The Kasner-deSitter spacetimes, like the Kasner spacetimes themselves, generally have a curvature singularity singularity at $t=0$.  The exception is the case that a single of the exponents {\it e.g.} $p_1=1$ with $p_k=0$ for $k=2,\dots,n$ which has a smooth early time limit.  Although the Kasner solution with these exponents is actually flat spacetime, the corresponding Kasner-Desitter spacetime is not diffeomorphic to deSitter.

We are accustomed to exponentially decaying terms not contributing to ADM charges in asymptotically flat spacetimes.  This will not be the case for asymptotically future deSitter spacetimes, because the spatial volume element is also growing exponentially with time.
However, it is possible to make the formulas look more familiar by converting to a new time coordinate such that the falloff to the asymptotic deSitter regime has a power law form.
This is done by setting
\begin{equation}
T=T_0\cosh^{2/n}({nt\over 2l})
\end{equation}
which brings the Kasner-deSitter spacetimes into a `brane-like' form
\begin{equation}\label{kasner-ds}
ds^2 = - {L^2\over T^2} F^{-1}(T) dT^2 +{T^2\over L^2}\left\{ \sum_{k=1}^n F^{p_k}(T)(dx^k)^2\right\}, \qquad 
F(T) = 1-{T_0^n\over T^n}
\end{equation}
The early time Kasner limit now happens in the limit that ${T/ T_0}=1+\epsilon$ with $\epsilon\ll 1$, and there is now a power law approach to the deSitter metric as $T/T_0$ becomes large.

For the discussion of cosmic hair, it is helpful to recognize the Kasner-dS solutions as analytic continuations of planar AdS solutions.
The mass and spatial tensions of these asymptotically AdS solutions can be computed using standard formulas, and will continue back to the computation of cosmological tension in the asymptotically dS case. 
Starting with the Kasner-dS spacetimes (\ref{kasner-ds}), one may analytically continue by setting $T=ir$, $T_0=ir_0$, $x^n=it$, $l=iL$ and relabel $p_n=p_0$ and arrive at
\begin{equation}\label{kasner-ads}
ds^2 =  {L^2\over r^2} F^{-1}(r) dr^2 +{r^2\over L^2}\left\{ - F^{p_0}(r) dt^2 +\sum_{k=1}^{n-1} F^{p_k}(r)(dx^k)^2\right\}
\end{equation}
where $F(r)= 1-r_0^n/r^n$ and the cosmological constant is now given by $\Lambda = - n(n-1)/2L^2$.  
The Kasner constraints are now given by
\begin{equation}\label{constraints}
p_0+ \sum_{k=1}^{n-1} p_k = 1 =p_0^2+\sum_{k=1}^{n-1} (p_k)^2
\end{equation}
As $r/r_0$ becomes large, these solutions now approach an AdS limit.  In the limit that $r/r_0=1+\epsilon$ with $\epsilon\ll 1$, these spacetimes approach the vacuum Levi-Civita solutions, which are analytic continuations of the Kasner spacetimes (\ref{kasner}), and for this reason we call them Levi-Civita-AdS spacetimes.
Certain special cases are well known.  Taking $p_0=1$ and $p_k=0$ for all $k$, gives the planar AdS black hole.  In this case $r=r_0$ is a smooth event horizon.  
Taking, for example, $p_1=1$ with $p_0=p_k=0$ for $k=2,\dots,n-1$ gives an AdS soliton \cite{Horowitz:1998ha}.  In this case, if the period of $x^1$ is chosen appropriately, then the space closes off smoothly at $r=r_0$.  These are the only examples such that $r=r_0$ is non-singular.  A further particular example of a spacetime of the form (\ref{kasner-ads}) was considered \cite{Myers:1999psa}, while the general case has been studied in \cite{Sarioglu:2009vq,Ren:2016xhb}.

The ADM mass and spatial tensions of Levi-Civita-AdS spacetimes (\ref{kasner-ads}) can be found using results of \cite{El-Menoufi:2013pza} and are given by
\begin{equation}\label{lcads}
{\cal M} =  { v r_0^n \over 16\pi l^D}(np_0-1),\qquad 
{\cal T}_k =  { v r_0^n \over 16\pi L_k l^{(n+1})}(np_k-1)
\end{equation}
where we have taken each of the spatial coordinates to be identified according to $x^k\equiv x^k+ L_k$ and defined the volume $v=\prod_{k=1}^{n-1} L_k$.  The mass and tensions of all asymptotically planar AdS solutions, including the Levi-Civita-AdS spacetimes, satisfy the sum rule \cite{El-Menoufi:2013pza,El-Menoufi:2013tca}
\begin{equation}\label{sumrule}
{\cal M}+\sum_{k=1}^{n-1}{\cal T}_k  L_k = 0
\end{equation}
which is closely related to tracelessness of the stress-tensor of the boundary CFT.   We should expect to find a similar sum rule for the cosmological tensions of asymptotically deSitter spacetimes.

\section{ADM cosmological tensions}\label{admsection}

The cosmological tensions for Kasner-deSitter spacetimes can be obtained via analytic continuation from the expression (\ref{lcads}) for the spatial tensions of the Levi-Civita-anti-deSitter spacetimes.  However, we will give a general construction that has wider applicability.
An ADM charge corresponding to an asymptotic symmetry of a spacetime is defined via Hamiltonian perturbation theory \cite{Regge:1974zd}. We follow the general prescription presented in \cite{Traschen:2001pb}, which we briefly outline here. The construction starts with foliating the spacetime by $(D-1)$-dimensional hypersurfaces $\Sigma$ with unit normal $n^a$ such that the metric can be decomposed as 
\begin{align*}
g_{ab}= (n\cdot n) n_an_b+s_{ab}
\end{align*}
where $n_a n^a =\pm 1$ and $s_{ab}$ denotes the induced metric on the slice(s), satisfying the orthogonality relation $s_{ab}n^b=0$.
Let $(s_{ab},\pi^{ab})$ denote the Hamiltonian initial data on a slice $\Sigma$ and $(h_{ab},p^{ab})$ be perturbations linearized about the background denoted by $(\bar{s}_{ab},\bar{\pi}^{ab})$. Furthermore, let $\xi^a$ be a Killing vector of the background, which we project along the slice $\Sigma$ and its normal according to $\xi^a=Fn^a+\beta^a$ such that $n_a\beta^a=0$.

The ADM charge corresponding to the Killing vector $\xi^a$ is then defined by an integral over a $(D-2)$-dimensional boundary at infinity on $\Sigma$ given by
\begin{align}{\label{admcharge}}
Q(\xi)=-\dfrac{1}{16\pi G}\int_{\partial\Sigma_{\infty}}da_cB^c
\end{align}
where
\begin{align}
B^a=F(\bar{D}^ah-\bar{D}_bh^{ab})-h\bar{D}^aF+h^{ab}\bar{D}_bF+\dfrac{1}{\sqrt{s}}\beta^b(\bar{\pi}^{cd}h_{cd}\bar{s}^a{}_b-2\bar{\pi}^{ac}h_{bc}-2p^a{}_b)
\end{align}
and $\bar{D}_a$ is the covariant derivative operator compatible with the background metric $\bar{s}_{ab}$. 

We now use this construction to obtain an expression for the cosmological tensions of an asymptotically future deSitter spacetime in terms of falloff coefficients.  
By asymptotically future deSitter we will mean that  at late times the metric has the form
\begin{align}\label{asyds}
ds^2\simeq-\dfrac{l^2}{T^2}\left(1+\dfrac{c_T}{T^n}\right)dT^2+\dfrac{T^2}{l^2}\sum_{k=1}^{n}\left(1+\dfrac{c_k}{T^n}\right){(dx^k)}^2
\end{align}
where $l$
is the de Sitter radius defined in (\ref{desitter}), $n=D-1$, and we neglect terms that fall off more rapidly.   The Kasner-deSitter spacetimes (\ref{kasner-ds}) have this form asymptotically.
Our subsequent results rely on the asymptotic fall off specified here. For example, if a metric approaches deSitter more slowly the tensions would diverge
as $T$ goes to infinity. 
 Note that there is a `gauge transformation' that leaves the asymptotic form (\ref{asyds}) invariant.  If we reparameterize the 
time coordinate according to 
\begin{equation}
T=\tilde T +{\lambda\over {\tilde T}^{n-1}}
\end{equation}
then the falloff constants shift according to
\begin{equation}
\tilde c_{\tilde T} = c_T-2n\lambda,\qquad  \tilde c_k = c_k+2\lambda
\end{equation}
The formula we arrive at for the cosmological tension charges below will be invariant under such a shift and therefore represents a `gauge invariant' measure of the falloff of an asymptotically future deSitter spacetime.

To obtain the cosmological tension along the spatial direction $x^k$, 
we take $\xi$ to be the Killing vector $\partial/\partial x^k$ of the deSitter background and 
we foliate the spacetime by constant-$x^k$ slices with unit normal $n=(T/l)dx^k$.  This determines the quantities $F=T/l$ and $\beta^a=0$. Finally, the perturbations to the metric on $\Sigma$ that follow from the asymptotic form of the spacetime metric (\ref{asyds}) are given in terms of the falloff coefficients $c_T$ and $c_k$ by
\be
h_{TT}=-\dfrac{l^2c_T}{T^{n+2}}, \qquad h_{ii}=\dfrac{c_i}{l^2T^{n-2}}, \qquad h=\dfrac{1}{T^n}\left(c_T+\sum_{i\neq k}c_i\right).
\ee
while the perturbation to the momentum vanishes.
We can now evaluate the integral (\ref{admcharge}) on the boundary at future infinity, resulting in the expression for the cosmological tension 
\be\label{ten}
\mathcal{T}_k= - 
\dfrac{v }{{16\pi L}_k l^{D}}\left(c_T  +n c_k\right).
\ee
where, as above, $ L_k$ is the range of $x^k$, $0\leq x^k \leq L_k$, and $v  =\prod_{k=1}^n L_k$. 
This expression can be processed further by virtue of the following.
For asymptotically AdS spacetimes, it was shown in \cite{El-Menoufi:2013pza} that the trace of the perturbation to the metric on 
$\Sigma$ vanishes in the asymptotic regime due to the equations of motion, which is a reflection of the sum rule (\ref{sumrule}), and is
related to the vanishing of the trace of the stress-tensor of the boundary CFT in AdS. An analogous 
result holds for asymptotically deSitter spacetimes\footnote{It is easily checked that the Kasner-deSitter spacetimes (\ref{kasner-ds}) have this property.},
that is, the cosmological tensions (\ref{tentwo}) satisfy a tracelessness relation\footnote{This condition should be related to a similar property of the trace of the boundary stress tensor in dS/CFT.}
\begin{align}\label{vanish}
\sum_{k=1}^n \mathcal{T}_k L_k=0
\end{align}
Hence the trace of the metric perturbation is zero in the far field. In terms of the fall-off coefficients in equation (\ref{asyds}) this implies that
\be\label{csum}
c_T + \sum_j c_j = 0
\ee
Therefore the $n$ tensions can be expressed in terms of just $n$, rather than $n+1$, falloff coefficients. Eliminating $c_T$ from the expression (\ref{ten}) gives
\be\label{tentwo} 
\mathcal{T}_k=\dfrac{ v }{16\pi L_k L^{D}}\left( -(n-1)c_k +\ \sum_{i\neq k}c_i\right)
\ee
For Kasner-deSitter spacetimes (\ref{kasner-ds}) the falloff coefficients are given by $c_T=T_0^n$ and $c_i=-p_iT_0^n$. Plugging these in to either expression (\ref{ten}) or (\ref{tentwo}) for the cosmological tension, and making use of the constraint (\ref{constraints}), yields the cosmological tensions for Kasner-deSitter spacetime 
\be\label{admtension}
\mathcal{T}_k=\dfrac{ v  T_0^n}{16\pi L_k L^{D}}(np_k-1)
\ee

The formula (\ref{tentwo}) for the cosmological tensions of a general asymptotically future deSitter spacetime and (\ref{admtension}) which gives cosmological tensions specifically for Kasner-deSitter spacetimes are among our main results.  They demonstrate that evidence of early time anisotropies can persist into the asymptotic deSitter regime in the form of ADM charges given by boundary integrals at future infinity, which constitute a form of cosmic hair.
The relation via analytic continuation to the mass and spatial tensions of asymptotically planar AdS spacetimes shows that cosmic hair is on a similar footing to more familiar types of hair that are defined via integrals evaluated at spatial infinity.

\section{Komar cosmological tensions and Smarr relations}\label{komarsection}

\subsection{General cosmological Smarr relation}
A Komar version of cosmological tension can also be defined if a spacetime has  a spatial translation symmetry throughout its evolution.  The existence of Komar tensions is thus less general, in the sense that the ADM construction above requires the existence of a spatial Killing vector only asymptotically in the late time regime.  However,  in another sense Komar cosmological tension is more widely applicable, since it can be defined without reference to any specific future asymptotic behavior.  For example, although we will primarily focus on AFdS spacetimes, Komar tensions can be defined for FRW cosmologies or vacuum Kasner solutions, which are not AFdS.

The construction of Komar charges and the associated Smarr relations starts with the differential identity for a Killing vector $\xi^a$
 \be\label{komar}
 \nabla_b\nabla^b\xi^a=  - R^a{}_b \xi^b
 \ee
Let  $\Sigma$ be a codimension-$1$ hypersurface in the spacetime with unit normal $n_a$
 and boundary $\partial\Sigma$. Using Stokes' theorem\footnote{For an anti-symmetric tensor $\alpha^{ab}=\alpha^{[ab]}$, Stokes theorem states that
$\int_{\partial\Sigma} ds_{ab}\alpha^{ab} = \int_\Sigma ds_b \nabla_a \alpha^{ab}$. } 
 equation (\ref{komar}) can then be rewritten as a `Komar integral relation' 
\begin{equation}\label{gauss}
\int_{\partial\Sigma}ds_{cb}\nabla^c \xi^b = - \int_\Sigma ds_c R^c {}_b\xi^b
\end{equation}
Here  $ds_c = da n_c$, where  $n_c $ is the normal to $\Sigma$,
and $ds_{cb}= da m_{[c} n_{b]} $, with $m_c$ the normal to $\partial\Sigma$,  and $da$ is the natural volume element on the appropriate
submanifold.  In equation (\ref{gauss}) Stokes Theorem requires that if $m_c$ is spacelike, then it is outward pointing from $\Sigma$, and
if $m_c$ is timelike it points into $\Sigma$.

Taking the hypersurface $\Sigma$ to extend out to a boundary, $\partial\Sigma_\infty$, at infinity (either spacelike or timelike), 
  the Komar charge $K$ associated with the Killing vector $\xi^a$ is then defined as the integral over this component of the boundary,
\begin{equation}\label{kchargedef}
K_{(\xi )}^{\infty} = - \int_{\partial\Sigma_\infty}ds_{cb} \nabla^c\xi^b
\end{equation}
If the hypersurface $\Sigma$ also has a boundary in the interior of the spacetime, then (\ref{gauss}) gives a relation, called a Smarr relation, between the Komar charge and the internal boundary integral, as well as a possible volume contribution if $T_{ab}\neq 0$. 

For example, consider a static asymptotically flat vacuum black hole and take $\Sigma$ to be a spacelike hypersurface that stretches between the black hole horizon and spatial infinity.  In this case, the Komar charge associated with the time translation Killing vector is proportional to the ADM mass.  
Equation  (\ref{gauss}) then shows that, with an appropriate choice of orientations, the boundary integral at infinity is equal to the boundary integral evaluated at the horizon, which is found to be proportional to $\kappa A$, with $\kappa $ the surface gravity and $A$ the horizon area.  This yields the Smarr relation
\begin{equation}\label{smarr1}
 (D-3) M = (D-2){\kappa A\over 8\pi}
\end{equation}
The Smarr relation can also be derived from the first law 
via a scaling argument, see {\it e.g.} \cite{Kastor:2009wy}.  In this one sees that the factors of $D-3$ and $D-2$ in (\ref{smarr1}) arise from the dimensions of the mass and horizon area respectively.

Now suppose that we have a cosmological spacetime with an exact spatial translation Killing vector $\xi^a$.  The spacetime may have a number of spatial translation Killing vectors, but at present we will focus on a single one,
 $\xi=\partial / \partial y ,$ associated with translation invariance in the $y$-direction.  We take the hypersurface $\Sigma$ to be timelike with normal in the $y$-direction, so that the unit normal to $\Sigma$ is given by $n_a = y_a$, where $y_a=F\nabla_a y$ and $F$ is a normalizing function.
We will assume that all the spatial directions  are compact and that the hypersurface $\Sigma$ wraps all the spatial directions except the $y$-direction, so that it has  boundaries only at an initial time $t_i$ and a final time $t_f$.  
In this context, we define the Komar cosmological tension $K_{(y)}(t )$ associated with the Killing vector $\xi =(\partial/\partial y )$
as the same boundary integral in the Komar charge $K_{(y)}^\infty $, equation (\ref{kchargedef}), 
but with the boundary taken at any time $t$, that is,
\begin{equation}\label{kcharge}
K_{(y)} (t  )=  \int_t   da\ t_c y_b  \nabla^c \xi^b
\end{equation}
Here $t^c$ is the future pointing timelike normal to $\partial \Sigma_t$, and the quantities  $m_c =- t_c$ and
 $ds_{cb}= da\  m_{[c}  y_{b]} =
-da\  t_{[c}  y_{b]} $ have been substituted into (\ref{kchargedef}).

Our way of proceeding here is slightly different from the asymptotically flat case.  In that case,  a Komar charge (\ref{kchargedef}) was defined only at infinite radius,  while a possible interior boundary term at a black hole horizon was evaluated in terms of explicit properties of the horizon.  In the cosmological case, there is not a clear analogue of a black hole horizon, and at this point in the construction we simply consider the early and late time boundary terms to be on the same footing, using
 $J=(i,f)$ in (\ref{kcharge})  to denote the initial and final boundaries, $\partial\Sigma_{J}$.
Later on, we will take the limit that $t_f$ goes to infinity.  However, it will be helpful to start by keeping $t_f$ general at this point.  Note that for the Kasner-deSitter spacetimes (\ref{kasner-ds}), which serve as our main examples of AFdS spacetimes, it is natural to choose the initial time to be $t_i=T_0$.

In cosmology one is interested in stress energy coming from
gauge and scalar fields and/or cosmological fluids, in addition to a possible cosmological constant.  For  solutions to the Einstein equations, the Ricci tensor is then given by 
\begin{equation}
R^b{}_c ={2\Lambda\over D-2} \delta^b_c + 2\pi \left ( T^b{}_c - {T\over D-2 }  \delta^b_c\right) 
\end{equation}
where $T_{ab}$ denotes the stress-energy, excluding the contribution from the cosmological constant $\Lambda$.
The Gauss law relation (\ref{gauss}) then tells us that the early and late time Komar tensions are related by the cosmological Smarr formula
\begin{equation}\label{smarrcos}
K_{(y)} ( t_f)- K_{(y)}(t_i)
= \int_\Sigma  dt\, d^{n-1}x \sqrt{-s}\  y_b\,  \xi^c \left[ {2\Lambda \over D-2 } \delta^b_c + 
2\pi \left( T^b{}_c - {T\over D-2} \delta^b_c  \right) \right]
\end{equation}
where $\sqrt{-s}$ is the area element on $\Sigma$ and the integral over the time coordinate runs between the early and late time boundaries at $t_i$ and $t_f$.
The Smarr formula (\ref{smarrcos}) implies
that for vacuum cosmologies the Komar tensions are conserved, $K_{(y)} (t_f ) =K_{(y)} (t_i )$. 
In non-vacuum cosmologies, the Komar tensions will typically be time dependent.
For  asymptotically flat  black holes,  the Smarr formula (\ref{smarr1}) and its generalizations relate Komar charges at spatial infinity
to properties of the black hole horizon in the interior of the spacetime.    In the cosmological case, one instead has a relation between Komar charges at two different times, that includes possible contributions from matter fields in the intervening volume.

\subsection{Komar tensions in terms of cosmological expansion rates}\label{cosmo}

In this section, we restate the expression for the Komar cosmological tensions in more physical terms and compute them in a number of  examples.
We begin by showing that  $K_{(y)} (t )$ is the Hubble expansion in the $y$-direction at time $t$ integrated over the  boundary component $\partial \Sigma_t$.  For a class of interesting spacetimes, we will see that this integrated quantity remains finite even near an early time singularity,
where the expansion rates in different spatial directions may diverge or vanish. 
For AFdS spacetimes, on the other hand, which are our main interest in this paper, the Komar tensions $K(t)$ diverge at late times due to the rapid increase in the area element. In the Smarr relation (\ref{smarrcos}) this is also reflected in the divergent volume integral on the right hand side.   We will address this in section (\ref{smarrsection}) by showing how the Komar construction may be modified along the lines of \cite{Kastor:2009wy}, by relating the Komar tensions to the finite ADM cosmological tensions of section (\ref{admsection}), in order to obtain finite versions of the Komar cosmological tensions in the AFdS case.

Our first task here  is to show that the Komar charge $K _{(y)}(t) $ defined in (\ref{kcharge}) can be rewritten in terms 
of the extrinsic curvature of the boundary component $\partial \Sigma_t$. To this end, let $\Phi_t$ be a family of constant time surfaces with unit normal $t_a$.
The boundary $\partial\Sigma _t$ is then a surface of constant $y$ at time $t$. 
Let us now decompose the spacetime metric as
\begin{equation}
g_{ab} = -t_a t_b + \gamma_{ab}
\end{equation}
where $\gamma_a{}^b t_b=0$. The extrinsic curvature of $\Phi_t $ is then given by 
$ K_{ab} = \gamma_a{}^c \nabla _c t_b$, which is just the familiar extrinsic curvature of constant time surfaces in cosmology.  We assume that the slices of constant $t$ can be chosen such that they are orthogonal
to the  slices of constant $y$ and to the Killing field $\xi=\partial/\partial y$, so that\footnote{In the usual Smarr construction, for static, asymptotically flat spacetimes, the hypersurface 
$\Sigma$ is a constant time slice.  The boundary is a sphere at large radius with normal $r_a$, and the analogous assumptions are that $r_a t^a=r_a \xi ^a =0$. }
 $t_a y^a =0$ and $t_a \xi^a =0$.  
 The integrand in (\ref{kcharge}) for the Komar cosmological tension can now be rewritten as
\begin{align}\label{extrin}
 t^c y^b  \nabla _c \xi _b & = - y^b\left[ \nabla_b (t^c \xi_c )   -\xi^c \nabla_b t_c \right] \\
&= \,    y^b\, \xi^c K_{cb}
\end{align}
Hence the Komar charge $K_{(y)} (t)$ is simply the integral of the appropriate component of the extrinsic curvature, which gives the Hubble expansion rate in the $y$-direction integrated over the remaining spatial dimensions
\be\label{qtwo}
K_{(y)} (t) =  \int_{\partial\Sigma_t} d^{D-2}x \sqrt{ \gamma }  K_y{}^y 
\ee
This gives a simple geometrical interpretation of the Komar charge. Further,  the extrinsic curvature of  the constant time slice $\Phi_t$
is related to the gravitional momentum density  by  $K^{ab}=[ -\pi^{ab} + \gamma^{ab} \pi/(D-2) ] /\sqrt{ \gamma} $, and so
 the Komar cosmological tension associated with the translation symmetry $\partial /\partial y$  is essentially the integrated
gravitational momentum 
\be\label{kmomentum}
K_{(y)}(t) = \int _{\partial\Sigma_t} d^{D-2}x \left( -\pi^{ab} y_a y_b + {\pi \over D-2 } \right)
\ee

It follows for a homogeneous but possibly anisotropic cosmology, that the Komar cosmological tension $K_{(y)}(t)$  is  simply the integral, over the other spatial dimensions, of the Hubble expansion rate in the $y$-direction.
For example, diagonal Bianchi Type I metrics 
\begin{equation}\label{bianchimetric}
ds^2 = -dt^2 +\sum_{k=1}^n a_k(t)^2 (dx^k)^2
\end{equation}
have  spatial translation Killing vectors $\xi_{(k)}=\partial/\partial x^k$ associated with each of the spatial directions with $k=1,\dots ,n$. 
Spatially flat FRW cosmologies, where the scale factors in the different spatial directions are all equal, are included as a special case.  
Take the spatial directions to be compactified with $x^k\equiv x^k +  L_k$ and define the coordinate 
volume $v =\prod_{k=1}^n  L_k$. 
From (\ref{qtwo})  the Komar cosmological tension in the $k$th spatial direction is then given by 
\begin{equation}\label{qhubble}
K_k (t)= {  v \over  L_k} (\prod_{l=1}^n a_l(t))  H_k
 \end{equation}
 where $H_k=\dot a_k/a_k$ is the Hubble parameter in the $k$th direction, which gets multiplied by the volume of the
 constant time slice.
 
For the Kasner spacetimes, which are solutions to the vacuum Einstein equation of Bianchi Type I form and were presented above in section (\ref{kasnersec}), the constraint (\ref{kasnerconstraints}) implies that the product of the scale factors is given by 
$\prod_{k=1}^n a_k(t)={t\over t_0}$
and the Komar cosmological tensions work out to be
 \begin{equation}
 K_k =  { v\,  p_k\over  t_0\, L_k  }\, 
 \end{equation}
proportional to the scale factor exponents in the corresponding direction and constant in accordance with the cosmological Smarr formula (\ref{smarrcos}) for vacuum spacetimes.  The Komar tensions for Kasner spacetimes also satisfy the sum rule $\sum_{j=1}^n  L_j K_j =  v/t_0$
as a consequence of the linear Kasner constraint (\ref{kasnerconstraints}).

As a further example, we consider $4D$ spatially flat FRW cosmologies, with all the scale factors given by a common function $a(t)$.  
Let us also assume that the compactification lengths $ L_k$ are all equal to a common value $ L$.   If we focus on power law cosmologies with $a(t) = (t/t_0)^q$,  then the corresponding perfect fluid matter has equation of state $p=w\rho$ with $w= {2\over 3q}-1$,  and the Komar tensions in the three spatial dimensions are all equal to
\begin{equation}
K(t) = {q L^2\over t_0}(t/t_0)^{3q-1} 
\end{equation}
Note that for $q=1/3$, corresponding to the equation of state parameter $w=1$, the Komar cosmological tension is independent of time\footnote{In this $q=1/3$ case, although the full tensor $T_{ab}-{1\over 2}g_{ab}T$ that appears on the right hand side of the cosmological Smarr formula (\ref{smarrcos}) for $D=4$ does not vanish, the relevant component does.}.  For exponents $q>1/3$ , the Komar cosmological tension vanishes at $t=0$ and diverges at late times, while for $q<1/3$ one finds the opposite behavior.  This infinite jump in the tension between the early and late time limits is sensible because the matter contribution on right hand side of the cosmological Smarr relation (\ref{smarrcos}) is integrated over the entire history of the universe.

A further special case is deSitter spacetime, which has a metric of the Bianchi Type I form (\ref{bianchimetric}) with scale factors all equal to $a(t)= \exp(2Ht)$, with the Hubble constant given by $H^2=2\Lambda/n(n-1)$. If we again take all the compactification lengths to be equal to a common value $L$ 
then the Komar cosmological tensions are again all equal and given by
\begin{equation}
K(t) = HL^2e^{nHt}
\end{equation}
We see that the Komar cosmological tension for deSitter diverges as $t\rightarrow \infty$, reflecting the integration on the right hand side of the cosmological Smarr formula (\ref{smarrcos}) of $\Lambda$ over an infinite spatial volume.  Since deSitter spacetime is trivially asymptotically future deSitter, with vanishing falloff coefficients, the ADM cosmological tension defined in section (\ref{admsection}) vanishes, and  we  have a clear conflict between the Komar and ADM definitions of cosmological tension, which both apply in this case.
This conflict will be resolved in the next section.

\section{Smarr relation for AFdS spacetimes}\label{smarrsection}

We now have two different notions of cosmological tension, which are defined for two different, but overlapping, classes of spacetimes.  The ADM cosmological tension of section (\ref{admsection}) is defined for AFdS spacetimes.  This includes inhomogeneous cosmologies, which have spatial translation symmetry only asymptotically at future infinity.  The Komar cosmological tension of section (\ref{komarsection}), on the other hand, which requires a spatial Killing vector throughout, is defined without reference to any specific future asymptotic behavior.
Both types of tension charges are defined for homogeneous, AFdS spacetimes, such as the Kasner-deSitter spacetimes (\ref{kasner-ds}).  However, as noted above the Komar cosmological tension for an AFdS spacetime diverges in the late time limit.  In this section, we will show how to fix this so that a new, modified Komar cosmological tension will agree with the ADM cosmological tension and yield a meaningful Smarr relation for this class of cosmological spacetimes.

\subsection{ADM mass vs. Komar mass and thermodynamic volume}
We start this disussion by taking a step backwards and considering static asymptotically flat spacetimes.
In this case, one finds that although the ADM and Komar masses are determined by different combinations of falloff coefficients, they nonetheless equal on solutions to the vacuum Einstein equations.  With a nonzero cosmological constant, this is no longer the case.  The ADM mass for spacetimes that are asymptotically (A)dS at spatial infinity is a finite quantity, while the expression for the Komar mass diverges. 
In \cite{Kastor:2009wy} we showed how the definition of the Komar construction could be modified when $\Lambda\neq 0$, in order to  remove this divergence and restore equality between the ADM and Komar masses.   
In this construction, the divergent part of the `bare' Komar charge is subtracted out, leaving behind a `renormalized' Komar mass that is proportional to the ADM mass, while the  subtracted piece combines with the infinite volume contribution to give a finite remainder.
The remainder appears as an additional term in the Smarr formula, which is now given by
\begin{equation}\label{smarrads}
(D-3) M=  (D-2){\kappa A\over 8\pi} -2{\Theta\Lambda\over 8\pi}
\end{equation}
The new quantity $\Theta$, which has dimensions of volume, is thermodynamically conjugate to $\Lambda$.  It also appears in the first law for asymptotically (A)dS black holes extended to include variations in the cosmological contant \cite{Kastor:2009wy}
\begin{equation}\label{firstads}
dM = {\kappa\over 8\pi} dA +{\Theta\over 8\pi} d\Lambda
\end{equation}
The Smarr formula (\ref{smarrads}) can also be derived from the first law (\ref{firstads}) via a scaling argument, whereby the factor of $-2$ in front of the last term arises as the scaling dimension of the cosmological constant.

An alternative formulation, which differs only slightly but assists in the interpretation of the new terms, is to view the cosmological constant as a (negative) pressure $P=-\Lambda/8\pi$ and to define the conjugate thermodynamic volume $V=-\Theta$.  
For a Schwarschild-AdS black hole with a spherical horizon, the thermodynamic volume is found to be
\begin{equation} 
V= {\Omega_{D-2} r_H^{D-1}\over D-1}
\end{equation}
where $r_H$ is the horizon radius and $\Omega_{n}$ is the area of a unit $n$-sphere.  It coincides with a naive computation of the volume inside the black hole horizon computed using the full $D$ dimensional volume element.  With this relabeling of variables, the first law (\ref{firstads}) now reads
\begin{equation}
dM = {\kappa\over 8\pi} dA + VdP
\end{equation}
One sees that the AdS black hole mass should be interpreted as an enthalpy, which differs from the internal energy by the amount of energy needed to create the system, {\it i.e.} replace the AdS vacuum by the black hole horizon and its interior.
See \cite{Cvetic:2010jb} for computation of the thermodynamic volume for spinning and charged AdS black holes with spherical event horizons.  See \cite{El-Menoufi:2013pza} for results on planar AdS black holes.

\subsection{Renormalized Komar mass in AFdS spacetimes} 
Returning to the cosmological case, the divergence of the Komar cosmological tension for AFdS spacetimes, which we seek to address, has essentially the same origin as the divergence in the `bare' Komar mass for spacetimes that are asymptotically (A)dS at spatial infinity,  and the construction of \cite{Kastor:2009wy} can be adapted to handle the AFdS case as well.
Assume that we have an AFdS spacetime with exact spatial translation symmetries, so that both the ADM and Komar definitions of cosmological tension are applicable.
Following \cite{Bazanski:1990qd,Kastor:2008xb,Kastor:2009wy}, we define a Killing potential $\omega^{ab}$ associated with a spatial translation Killing vector $\xi^a$ to be an antisymmetric tensor, $\omega^{ab}=\omega^{[ab]}$, satisfying
\be\label{kp}
\nabla_c \omega^{ca} = \xi^a
\ee
The existence of a Killing potential, at least locally, follows from Poincare's lemma, since the divergence of a Killing vector vanishes\footnote{The Killing potential $\omega^{ab}$ is not uniquely determined by (\ref{kp}).  If $\alpha^{ab}$ is an antisymmetric tensor satisfying $\nabla_a\alpha^{ab}=0$, then $\tilde\omega^{ab}=\omega^{ab}+\alpha^{ab}$ also solves (\ref{kp}).  However, this non-uniqueness does not impact our results.}.

Now consider the cosmological Smarr relation (\ref{smarrcos}) in the AFdS context.  To simplify the formulas, we will take the stress energy tensor to vanish, leaving only the contribution from $\Lambda>0$ on the right hand side, so that
\begin{equation}\label{again}
K_{(y)} ( t_f)- K_{(y)} (t_i)
= {2\Lambda \over D-2 }\int_\Sigma  dt\, d^{n-1}x \sqrt{-s}\  y_a\,  \xi^a 
\end{equation}
where, as before, the timelike hypersurface $\Sigma$ stretches between boundaries at an initial time $t_i$ and a final time $t_f$.
If in order to make a comparison with the ADM tension, we let $t_f$ tend to infinity, then $K_{(y)}( t_f)$ will diverge due to the infinite volume of integration on the right hand side of (\ref{again}).  

The Killing potential $\omega^{ab}$, introduced above, can be used to rewrite the volume integral on the right hand side of (\ref{again}) as a boundary integral.  It follows from (\ref{kp}) that $ y_a \xi^a = D_a( \omega ^{ab}y_b) $, where $D_a$ is the covariant derivative operator compatible with the metric $s_{ab}$ on the hypersurface $\Sigma$.  Recalling the definition (\ref{kcharge}) of $K_{(y)}(t)$, we then rewrite the cosmological Smarr relation  (\ref{again}) in the form
\begin{equation}\label{newkomar}
 \int_{\partial\Sigma_{f}} da y_bt_c  \left( \nabla ^b \xi ^c  +{2\Lambda\over D-2}  \omega^{bc} \right) =  \int_{\partial\Sigma_{i}} da y_b t_c \left( \nabla ^b \xi ^c  +{2\Lambda\over D-2}  \omega^{bc} \right)
\end{equation}
and observe that there is again a conserved quantity, as was the case for the original Komar cosmological tension in the vacuum case.  Let us now address the issue of divergences by looking at the quantities $\nabla^a\xi^b$ and $\omega^{ab}$ in the asymptotic regime (\ref{asyds}).  One finds that  the non-zero components of these quantities in this limit, up to antisymmetry, are given by
\begin{align}\label{asymp1}
\nabla^T\xi^k & \simeq -{T\over l^2} +{1\over l^2T^{n-1}}({n\over 2} c_k +c_T)\\
\omega^{Tk}&\simeq +{T\over n} +{\alpha\over T^{n-1}} \label{asymp2}
\end{align}
where we have now identified $y \equiv x^k$.  Here $\alpha$ is an arbitrary constant, reflecting the possibility of adding a homogeneous term to a particular solution for the Killing potental (\ref{kp}).  One finds that if this constant is taken to have the value 
$\alpha = -c_T/2n$, then the quantity on the left hand side of (\ref{newkomar}) will be finite and proportional to the ADM cosmological tension (\ref{ten}) as the final time $t_f$ is taken to infinity.  Let us therefore define the quantity 
$\omega_{div}^{ab}=\omega_{div}^{[ab]}$, which diverges in the late time limit, to have the non-zero components
\begin{equation}\label{omegadiv}
\omega_{div}^{Tk}={T\over n} - {c_T\over 2n T^{n-1}}
\end{equation}
and also define the renormalized Komar cosmological tension by
\begin{equation}
K_{(y)}^{ren} = K_{(y)} +  {2\Lambda\over D-2}  \int_{\partial\Sigma_{\infty}} da\ t_c  y_b  \omega_{div}^{cb}
\end{equation}
where we have taken the limit that the final time $t_f$ goes to infinity.    
Making use of the asymptotic forms (\ref{asymp1}) and (\ref{asymp2}) and comparing with the expression (\ref{ten}) for the ADM tension ${\cal T}_k$, we see that
\begin{equation}
K_{(y)}^{ren} =  8\pi{\mathcal T}_{(y)}
\end{equation}
Hence the cosmological Smarr formula (\ref{newkomar}) can now be rewritten in terms of the tension ${\cal T}_{(y)}$ by adding in and subtractiong
out $ \omega_{div}^{cb}$ in the integration over the late time boundary $\partial\Sigma_{\infty}$ to give the Smarr formula in terms of 
the cosmological tension. Changing the index `$(y)$' to the index `$(k)$' on the cosmological tension and associated quantities (in accordance with the identification of $y\equiv x^k$ made above) the cosmological Smarr formula can now be written as\footnote{Including a cosmological fluid stress energy, the result would be
\be\label{smarr3}
 8\pi{\mathcal T}_{(k)} =  K_{(k)} (t_i )   - {2\Lambda\over D-2} V_{(k)} +
4\pi \int_\Sigma \sqrt{-g} (\rho + (n-2) p_k- \sum_{i\neq k} p_i )
\ee
}
\begin{equation}\label{finalsmarr}
 8\pi{\mathcal T}_{(k)}  = K_{(k)} (t_i) - {2\Lambda\over D-2} V_{(k)}
\end{equation}
where the definition of the cosmological tension is given in equation (\ref{admcharge}) or (\ref{ten}),
and that of the  intial time boundary contribution in (\ref{kcharge}) or (\ref{qtwo}).
The new quantity $V_{(k)}$ in this expression, which is the analogue of the thermodynamic volume in the AFdS case and will be called the ``cosmological volume", is given by
\begin{equation}\label{voldef}
V_{(k)} =  \int_{\partial\Sigma_\infty}da_{cb} (\omega^{cb}-\omega^{cb}_{div}) -  \int_{\partial\Sigma_{i}} da_{cb} \omega^{cb} 
\end{equation}
The cosmological volume has dimensions $(length)^{D-1}$ and is rendered finite by the subtraction at infinity of the quantity $\omega^{ab}_{div}$. The cosmological volume $V_{(k)}$ depends on the choice of spatial Killing vector $\xi=\partial/\partial x^k$.  However, all the $V_{(k)}$ turn out to be related to a common value $V_{cosmo}$ with dimensions $(length)^D$ through
\begin{equation}
V_{(k)} = V_{cosmo} /L_k
\end{equation}
This follows because the Killing potentials corresponding to Killing vectors in different spatial translation symmetries ({\it e.g.} for $k=1,\dots,D-1$ if there is translational symmetry in all spatial directions) solve essentially the same equations, namely
\begin{align}
{1\over\sqrt{-g}}\partial_a\left(\sqrt{-g}\omega^{ak}\right)& = 1\\
{1\over\sqrt{-g}}\partial_a\left(\sqrt{-g}\omega^{al}\right)& = 0,\qquad l\neq k\nonumber
\end{align}
Differences in the cosmological volumes $V_{(k)}$ then reflect only differences in the ranges of integration.

The validity of the cosmological Smarr formula for AFdS spacetimes (\ref{finalsmarr}) can be checked by calculating all the relevant quantities in the Kasner-DeSitter spacetimes (\ref{kasner-ds}).  The ADM cosmological tensions, which appear on the left hand side, are given in (\ref{admtension}), while the early time Komar boundary term, taken to be at $t_i=T_0$, and the cosmological volume, which appear on the right hand side are found to be
\begin{equation}\label{check}
K_{(k)} (T_0) = -{v T_0^{D-1}\over 2L_k l^D} (D-1)p_k,\qquad 
V_{(k)} =  {v T_0^{D-1}\over 2 (D-1) L_k  l^{D-2} } 
\end{equation}
where we have taken the early time boundary to be $t_i=T_0$.
Substituting the definition of the deSitter length scale  $\Lambda = (D-1)(D-2)/2l^2$,
one sees that (\ref{finalsmarr}) is satisfied.

\subsection{Physical interpretation}

The Smarr relation for AFdS spacetimes (\ref{finalsmarr}) relates three quantities; the cosmological tension, which is defined at future infinity; an early time boundary term, which in case of Kasner-deSitter spacetimes it is natural to evaluate at the initial singularity $T_0$, and a cosmological volume term.  Let us discuss each of these in more physical terms, beginning with the tension, which plays an analogous role in (\ref{finalsmarr}) to the ADM mass in the Smarr formula for black holes (\ref{smarrads}).  Recall that in section (\ref{cosmo}) above, we related the Komar cosmological tension in the $y$-direction to an integral (\ref{qtwo}) of the component $K_y{}^y$ of the extrinsic curvature, which diverges at late times for AFdS spacetimes.   Accordingly, it is the renormalized Komar tension, which is proportional to the ADM cosmological tension, that enters the Smarr relation.  If the metric in the asymptotic regime (\ref{asyds}) is rewritten in terms of proper time ({\it i.e.} with $g_{tt}=-1$), then we find that the renormalized Komar tension can be written as
\begin{equation}
K_{(y)}^{ren} = \int_{\partial\Sigma_{\infty}} d^{D-2}x\sqrt{\gamma}\left({{\dot a_y}\over a_y}-H_{dS}\right)
\end{equation}
where $H_{dS}=1/l$ is the Hubble constant of the asymptotic deSitter metric and $\sqrt{\gamma}$ is the exponentially growing volume element on $\partial\Sigma_t$ in the limit $t\rightarrow\infty$.  The instantaneous expansion rate ${\dot a_y}/a_y$ is approaching $H_{dS}$ in this limit, and the integral picks out the exponentially suppressed difference.  The cosmological tension boundary term at infinity, therefore, captures a trace of the anisotropy in the expansion rate as it dies off in the asymptotic limit.  This quantity would be measurable, in principle, to an observer living in such an asymptotic region.

The boundary term $K_{(y)} (T_0)$ is the integrated expansion rate in the $y$-direction,
\begin{equation}  
K_{(y)}(T_0) = \int_{\partial\Sigma_{T_0}} d^{D-2}x\sqrt{\gamma}\left({{\dot a_y}\over a_y}\right)
\end{equation}
but now evaluated at an early time, in the interior of the spacetime, away from the asymptotic limit.  It is the analogue of the 
$\kappa A$ boundary term that appears in the Smarr relations for black holes (\ref{smarrads}).  A key feature of this term is its finiteness, at least for Kasner-deSitter spacetimes (\ref{kasner-ds}), despite being evaluated in a singular limit of the spacetime\footnote{Outside of the AFdS context, this term is also finite at the initial singularities of Kasner spacetimes and FRW cosmologies with $a(t) \sim t^q$ for  $0<q<1/3$. }.  One finds that either the area element of a spatial slice
is going to zero with the expansion rate diverging, or the other way around in such a way that the product, which is essentially the gravitational momentum, stays finite.
The special case $p_1 = 1$, with $ p_j = 0$ otherwise, is an analytic continuation of
the AdS black hole metric, and one would therefore expect $K_{(j)} (T_0) $ to be finite in this case. However,
 it continues to be true for the whole Kasner-deSitter family.
 
The cosmological volume for Kasner-deSitter spacetimes, given in (\ref{check}), can be viewed as the spacetime volume ``before the big bang".   The quantity
\begin{equation}
V_{cos} =  {1\over 2}{v T_0^{D-1}\over  (D-1)  l^{D-2} } 
\end{equation}
 is equal to one-half times the volume of the deSitter 
background spacetime between $T=0$ and the big bang at $T_0$. This is also true more generally,
as is shown in the appendix.  This has the same flavor as the thermodynamic volume of AdS black holes, which appears in the Smarr formula (\ref{smarrads}) and represents the energy necessary to replace a portion of AdS spacetime with the black hole \cite{Kastor:2009wy}.
To summarize, the cosmological Smarr relation (\ref{finalsmarr}) relates these three quanities, the cosmological tension, the early time boundary term and the cosmological volume, in much the same way as the ADM mass, $\kappa A$ boundary term at the horizon and thermodynamic volume are related by the Smarr relation (\ref{smarrads}) for AdS black holes.

There are also interesting ways to combine the Smarr relations (\ref{finalsmarr}) and the tension sum rule  (\ref{vanish}). Summing over (\ref{finalsmarr}) for all the spatial directions and
using the sum reule (\ref{vanish}) gives an expression for  the cosmological volume in terms of the early time expansions 
\begin{equation}\label{earlysum}
 V_{cos} = {D-2  \over 2 (D-1) \Lambda }   \sum_{k=1}^{D-1}L_k  K_{(k)} ( T_0)  
\end{equation}
Note that (\ref{earlysum}) could be plugged back in to eliminate the cosmological volume from the Smarr relation.
Alternatively, if one takes the difference between the Smarr relations corresponding to two distinct spatial directions $j$ and $k$, weighted by factors of the respective compactification lengths,  then the cosmological volume terms drop out, yielding a relation between the cosmological tensions in these directions, which are evaluated at late times, and the early time boundary terms,
 \begin{equation}\label{strain}
 L_j  {\mathcal T}_{(j)} ( t_i)  - L_k  {\mathcal T}_{(k)} =  {1\over 8\pi}\left(L_j  K_{(j)} ( T_0)  - L_k  K_{(k)} ( T_0)\right)
\end{equation}
Since this involves the difference in tensions between different directions, we call this the `strain' relation.
As noted above, the ADM tensions are in principle measurable by an observer in the AFdS regime.
The strain relation (\ref{strain}) therefore demonstrates that information about the pre-inflationary phase of an AFdS universe, the right hand side of the equation, can be inferred from observable late time cosmic hair.

\section{Conclusion}\label{conclusion}

We have shown in this paper that AFdS spacetimes are characterized by a fundamental type of cosmic hair, called cosmological tension, which captures the leading order anisotropic corrections to the late time isotropic deSitter expansion rate.  As a consequence of the field equations in the asymptotic region, the cosmological tensions necessarily satisfy the sum rule (\ref{vanish}).  The cosmological tension charges are measurable for observers in the late time regime of an AFdS spacetime, if they are capable of accurately measuring the history of the expansion rates in the different directions.
We have also developed  Komar type cosmological tension charges, which are defined whenever a cosmological spacetime has exact spatial translation symmetries.  For AFdS spacetimes with translation symmetries this leads to a Smarr relation that relates cosmological tension to properties of the spacetime at early times, together with a cosmological analogue of the AdS black hole thermodynamic volume. 

There are a number of possible future directions for work in this area.
Cosmologists must deal with a messy universe, containing many kinds of matter fields, both known and unknown, and 
in inflationary models, the cosmological constant is often only approximately constant.  One important topic 
for future work is to analyze spacetimes that are asymptotically quasi-deSitter case, where {\it e.g.} 
 an approximate cosmological constant is generated by a scalar field potential.   `Cosmic no hair theorems' analogous to the results of \cite{Wald:1983ky} have been proved with these fields and asymptotic behaviors, establishing that large classes of such spacetimes exist \cite{Kitada:1991ih,Kitada:1992uh}.  It would be interesting to ask how fundamental ADM cosmic hair works in these systems? Whether the sum rule (\ref{vanish}) gets modified?  What is the analogue of cosmological volume?

Another issue for future work is to allow for perturbative inhomogeneities by deriving the analog of a first law for cosmological tension. It is not obvious
that there is a sensible perturbative formulation of the problem with the early time boundary term approaching a big bang singularity. However,
 the fact that the gravitational momentum was seen to be finite approaching the early time singularity in interesting examples, suggests that this might nonetheless be feasible. Taking into account these and other corrections - inhomogeneities and inclusion of matter fields - improvements to the idealized relations presented here could be used with observations of large scale anisotropies in the CMBR to further constrain models of inflation.

Finally, it would be interesting to see whether the realization that AFdS and potentially other classes of spacetimes are characterized by ADM charges can lead to progress in classifying cosmological spacetimes in analogy with well known results in black hole physics.  For example, is it true that the cosmological tensions uniquely specify Bianchi Type I solutions of the Einstein-$\Lambda$ field equations to be members of the Kasner-deSitter family of solutions. 

\subsection*{Acknowledgements}  The work of S.R. is supported by FONDECYT grant 1150907. He would also like to thank the Amherst Center for Fundamental Interactions at UMass, Amherst for their hospitality, and Adrian and Valerie Parsegian for the wonderful accommodations at the Casa F\'isica in Amherst during his visit when this work was initiated.

\vskip 0.2in

\appendix

{\huge\bf Appendix}

\section{Cosmological Volume}\label{cosmovolume}
For Kasner-deSitter spacetimes, it was noted that the cosmological volume 
$V_{cos} $ is simply related to the volume of spacetime ``before the big bang" singularity at $T=T_0$. In this appendix,  we will 
justify this statement and show that it holds for a wider class of spacetimes as well.
To this end, we first note that the spacetime volume element for Kasner-deSitter spacetimes in the coordinates of equation (\ref{kasner-ds}) is the same
 as that for deSitter. More generally, consider any metric of the form
 \begin{equation}\label{form}
ds^2 =- {l^2 \over \eta^2 } A(\eta ) d\eta^2  + {\eta^2 \over l^2} f_{ij}(\eta )dx^i dx^j 
\end{equation}
By transforming to a new time coordinate $T$, such that $dT / T = (d\eta /\eta) [ A(\eta ) det(f_{ij} (\eta ) ) ]^{-1/2}$,
the metric can be seen to have the property that   
\begin{equation}
\sqrt {-g} = \sqrt {-g_{dS} }  =(T/l)^{D-2} 
\end{equation}
so that it also has the same volume element as deSitter spacetime.
The integral of the quantity $\omega_{div}^{ab} $ on the
boundary at infinity can then be processed using the relations
\begin{align}
   \omega_{div}^{bc}  T_b y_c\,  da &= \sqrt{-g_{dS} }\  \omega_{div}^{Ty}\\
   & =\  { T^{D-2} \over l^{D-2}  }
    \left( {T\over  (D-1)}   + {\alpha \over  T^{D-2} }  \right) \nonumber
\end{align}
The contribution from the homogeneous solution with coefficient $\alpha$ is  independent of the time $T$.
 The non-constant term  can be converted to a volume term in deSitter using Stokes theorem, so that we have
\be\label{btvol}
\int_{\partial\Sigma_{T_f}} da\  T_b y_c \omega_{div}  ^{bc}  = {\alpha v\over l^{D-2} } -\int _{T=0}^{T_f} \  \sqrt{-g_{dS} } 
\ee
Hence the quantity $V_{(y)}$, defined in (\ref{voldef}) as the difference between the two boundary terms, becomes
\begin{align}\label{volumetwo}
V_{(y)}  =& -\int_{T_0}^{T_f} \sqrt{-g}\  +\  \int _{T=0}^{T_f} \  \sqrt{-g} \\  
 = &\int_{T=0}^{T_0} \sqrt{-g}  + {\alpha v\over l^{D-2} }  
\end{align}
and $V$ is seen to be the spacetime volume that is excluded between  $T=0$ and $T_0$, relative to pure deSitter spacetime, plus a constant from the choice of 
homogeneous solution.

\end{document}